\documentstyle[12pt]{article}
\begin{document}
\title{Further Effects of Varying $G$}
\author{B.G. Sidharth$^*$\\
Centre for Applicable Mathematics \& Computer Sciences\\
B.M. Birla Science Centre, Adarsh Nagar, Hyderabad - 500 063 (India)}
\date{}
\maketitle
\footnotetext{$^*$Email:birlasc@hd1.vsnl.net.in; birlard@ap.nic.in}
\begin{abstract}
The correct perihelion precession was recently deduced within the frame work
of a time varying Gravitational constant G. Here, we deduce also the
observed gravitational bending of light and flattening of galactic
rotational curves.
\end{abstract}
\section{Introduction}
In a recent communication\cite{r1} we saw that it is possible to account for
the precession of the perihelion of Mercury, for example, only in terms of
the time varying universal constant of gravitation G. It may be mentioned
that Dirac had argued\cite{r2} that a time varying G could be reconciled
with General Relativity and the perihelion precession by considering a suitable
redefinition of units. We will now show that it is also possible to account for the bending
of light on the one hand and on the other, the flat galactic rotation curves
without invoking dark matter, with the same time variation of G.\\
\section{Bending of Light}
It may also be mentioned that some varying G cosmologies have been reviewed
by Narlikar and Barrow, while a fluctuational cosmology with the above G
variation has been considered by the author \cite{r3,r4,r5,r6} and
\cite{r7}.\\
We start by observing that, as is well known, the bending of light
can be deduced in Newtonian theory also, though the amount of bending is half
of that predicted by General Relativity\cite{r8,r9,r10,r11}.
In this case the equations for the
orbit of a particle of mass $m$ are used in the limit $m \to 0$ with due
justification. A quick way
of obtaining the result is to observe that we have the well known orbital
equations\cite{r1,r12}.
\begin{equation}
\frac{1}{r} = \frac{GM}{L^2} (1+ecos\Theta)\label{e1}
\end{equation}
where $M$ is the mass of the central object, $L$ is the angular momentum per
unit mass, which in our case is $bc$, $b$ being the impact parameter or
minimum approach distance of light to the object, and $e$ the eccentricity
of the trajectory is given by
\begin{equation}
e^2 = 1+ \frac{c^2L^2}{G^2M^2}\label{e2}
\end{equation}
For the bending of light, if we substitute in (\ref{e1}), $r = \pm \infty$, and
then use (\ref{e2}) we get
\begin{equation}
\alpha = \frac{2GM}{bc^2}\label{e3}
\end{equation}
$\alpha$ being the deflection or bending of the light. This is half the General
Relativistic value.\\
We also note that the effect of time variation is given by (cf.ref.\cite{r1})
\begin{equation}
G = G_0 (1-\frac{t}{t_0}),r = r_0 (1-\frac{t}{t_0})\label{e4}
\end{equation}
where $t_0$ is the present age of the universe and $t$ is the time elapsed from
the present epoch.\\
Using (\ref{e4}) the well known equation for the trajectory is given by
(Cf.\cite{r13},\cite{r12},\cite{r14})
\begin{equation}
u" + u = \frac{GM}{L^2} + u\frac{t}{t_0} + 0 \left ( \frac{t}{t_0}\right )^2\label{e5}
\end{equation}
where $u = \frac{1}{r}$ and primes denote differenciation with respect to
$\Theta$.\\
The first term on the right hand side represents the Newtonian contribution
while the remaining terms are the contributions due to (\ref{e4}). The
solution of (\ref{e5}) is given by
\begin{equation}
u = \frac{GM}{L^2} \left[ 1 + ecos\left\{ \left(1-\frac{t}{2t_0}\right )
\Theta + \omega\right\}\right]\label{e6}
\end{equation}
where $\omega$ is a constant of integration. Corresponding to $-\infty < r < \infty$
in the Newtonian case we have in the present case, $-t_0 < t < t_0$, where
$t_0$ is large and infinite for practical purposes. Accordingly the analogue
of the reception of light for the observer, viz., $r = + \infty$ in the
Newtonian case is obtained by taking $t = t_0$ in (\ref{e6}) which gives
\begin{equation}
u = \frac{GM}{L^2} + ecos \left(\frac{\Theta}{2}
+ \omega \right)\label{e7}
\end{equation}
Comparison of (\ref{e7}) with the Newtonian solution obtained by neglecting
terms $\sim t/t_0$ in equations (\ref{e4}),(\ref{e5}) and (\ref{e6}) shows
that the Newtonian $\Theta$ is replaced by $\frac{\Theta}{2}$, whence the
deflection obtained by equating the left side of (\ref{e6}) or (\ref{e7})
to zero, is
\begin{equation}
cos \Theta \left(1-\frac{t}{2t_0}\right) = -\frac{1}{e}\label{e8}
\end{equation}
where $e$ is given by (\ref{e2}). The value of the deflection from
(\ref{e8}) is twice the Newtonian
deflection given by (\ref{e3}). That is the deflection $\alpha$ is
now given not by (\ref{e3}) but by
$$\alpha = \frac{4GM}{bc^2},$$
which is the correct General Relativistic Formula.
\section{Galactic Rotation}
The problem of galactic rotational curves is well known (cf.ref.\cite{r8}).
We would expect, on the basis of straightforward dynamics that the rotational
velocities at the edges of galaxies would fall off according to
\begin{equation}
v^2 \approx \frac{GM}{r}\label{e9}
\end{equation}
whereas it is found that the velocities tend to a constant value,
\begin{equation}
v \sim 300km/sec\label{e10}
\end{equation}
This has lead to the hypothesis of as yet undetected dark matter, that is that
the galaxies are more massive than their visible material content indicates.\\
We observe that from (\ref{e4}) it can be easily deduced that
\begin{equation}
a \equiv (\ddot{r}_{o} - \ddot{r}) \approx \frac{1}{t_o} (t\ddot{r_o} + 2\dot r_o)
\approx -2 \frac{r_o}{t^2_o}\label{e11}
\end{equation}
as we are considering infinitesimal intervals $t$ and nearly circular orbits.
Equation (\ref{e11}) shows (Cf.ref\cite{r1} also) that there is an anomalous inward acceleration, as
if there is an extra attractive force, or an additional central mass.\\
So,
\begin{equation}
\frac{GMm}{r^2} + \frac{2mr}{t^2_o} \approx \frac{mv^2}{r}\label{e12}
\end{equation}
From (\ref{e12}) it follows that
\begin{equation}
v \approx \left(\frac{2r^2}{t^2_o} + \frac{GM}{r}\right)^{1/2}
\label{e13}
\end{equation}
From (\ref{e13}) it is easily seen that at distances within the edge of a typical
galaxy, that is $r < 10^{23}cms$ the equation (\ref{e9}) holds but as we reach
the edge and beyond, that is for $r \geq 10^{24}cms$ we have $v \sim 10^7 cms$
per second, in agreement with (\ref{e10}).\\
Thus the time variation of G given in equation (\ref{e4}) explains observation
without taking recourse to dark matter.


\begin{thebibliography}{99}
\bibitem {r1} B.G. Sidharth, "Effects of Varying G" to appear in Nuovo Cimento B.
\bibitem {r2} P.A.M. Dirac, "Directions in Physics", Wiley-Interscience, New
York, 1978, p.79.
\bibitem {r3} J.V. Narlikar,  Foundations of Physics, Vol.13. No.3, 1983.
\bibitem {r4} J.D. Barrow and Paul Parsons, Physical Review D, Vol.55, No.4, 1997.
\bibitem {r5} B.G. Sidharth, Int.J.Mod.Phys.A, 13 (15), 1998, p.2599ff.
\bibitem {r6} B.G. Sidharth, Int.J.Th.Phys. 37(4), 1998, pp.1307ff.
\bibitem {r7} B.G. Sidharth, "Instantaneous Action at a distance in a holistic
universe", Invited submission to, "Instantaneous action at a distance in
Modern Physics: Pros and Contra", Eds., A. Chubykalo and R. Smirnov-Rueda,
"Nova Science Books and Journals", New York, 1999.
\bibitem {r8} J.V. Narlikar, "Introduction to Cosmology", Cambridge University
Press, Cambridge, 1993.
\bibitem {r9} H.H. Denman, Am.J.Phys. 51(1), 1983, 71.
\bibitem {r10} M.P. Silverman, Am.J.Phys. \underline{48}, 1980, 72.
\bibitem {r11} D.R. Brill and D. Goel, Am.J.Phys. 67(4), 1999, 317.
\bibitem {r12} H. Goldstein, "Classical Mechanics", Addison-Wesley, Reading,
Mass., 1966.
\bibitem {r13} P.G. Bergmann, "Introduction to the Theory of Relativity",
Prentice-Hall (New Delhi), 1969, p248ff.
\bibitem {r14} H. Lass, "Vector and Tensor Analysis", McGraw-Hill Book Co.,
Tokyo, 1950, p295 ff.
\end{thebibliography}
\end{document}